\documentclass[11pt,a4paper]{article}

\usepackage[margin=2.5cm]{geometry}
\usepackage{amsmath,amssymb}
\usepackage{amsthm}
\usepackage{booktabs}
\usepackage{longtable}
\usepackage{array}
\usepackage{graphicx}
\usepackage{caption}
\usepackage{enumitem}
\usepackage[hidelinks]{hyperref}

\newtheorem{proposition}{Proposition}
\newtheorem{corollary}{Corollary}

\title{\bfseries Sunk Slippage Is Not a General Cost:\\ A Structural Census of 901 DeFi Exploit Proofs-of-Concept}
\author{Sihao Kou \and Qingni Shen}
\date{November 2026}

\begin{document}
\maketitle

\begin{abstract}
\noindent
A cost model for DeFi attacks has to decide how to charge the price impact an attacker incurs
while moving a pool; treating it as a sunk cost is a natural default. Whether that default fits
depends on how attacks actually exit, and that distribution has not been measured. We reconstruct
the execution structure of 901 proof-of-concept exploits from DeFiHackLabs, spanning July 2017 to
October 2026, along three steps (where the capital is borrowed, where the price is moved, and
where the position is exited) and
report what a cost model has to be calibrated against.

Among the 149 price- and oracle-manipulation incidents in the corpus, 45.6\% move the price and
exit through the same pool (42.3\% on constant-product pools, 3.4\% on Curve stable pools); after
coding all 48 single-leg files by hand and recovering the round trips a static classifier missed,
the share rises to 51.0\%. Within the single-leg records, 10 (21.3\%) exit without returning to the
pool whose price they moved; this is the only structure in which sunk slippage actually exists.
Another 25 (53.2\%) move the price by direct writes (token transfers, \texttt{skim}, \texttt{sync},
\texttt{donate}) or by accounting-state inflation rather than by swaps, incurring almost no swap cost at all.

Sunk slippage is therefore a real cost in 6.7\% of price-manipulation incidents, not in
general. Applying the same settlement to the thirteen incidents in our parameterised set that
demonstrably paid, five (38\%) would be judged unprofitable, and every one of those reversals is
produced by the slippage term: fees and competition alone flip none. For most attacks the cost that
matters is gas, flash-loan fees and competition, and the accounting question is not how much slippage
to charge but whether the reported profit has already had each fee deducted. We release the structural taxonomy, a reproducible distribution, the static classifier, the coding
manual and three fork executions (one per structure). The classifier's agreement on a 30-case
manual audit is 93\% (28 of 30), and we report every known direction of error.

\vspace{0.6em}
\end{abstract}

\section{Introduction}

Automated exploit generation and vulnerability validation tools already decide, at the execution
level, whether a candidate exploit succeeds: A1 executes proofs-of-concept on forked state and
reports only those with positive net profit~\cite{a1}, and V2E treats triggerability and
profitability as joint criteria~\cite{v2e}. Their verdicts, however, are bound to the block state
that was tested. The natural next question (under which market conditions does that verdict
still hold?) has attracted a body of modelling work that adds execution cost, slippage and
competition terms to exploit profit.

Whatever form such a model takes, one input precedes the others: how the costs it charges are
distributed over real attacks. A term that dominates one execution structure can be irrelevant in
another, and the difference is mechanical rather than statistical. We therefore take no prior
position on whether any particular published model is wrong. What we supply is the distribution
such a model has to be calibrated against, and the constraint it places on what a slippage term
can mean.

Three observations organise the paper.

\emph{First}, round trips dominate. In 149 price- and oracle-manipulation incidents, 51.0\% move
the price and exit through the same pool. On a constant-product pool, the return leg walks the
curve back: the price impact is \emph{borrowed}, not spent. Only transaction fees are lost. For the
one case we reconstruct trade by trade, the true round-trip cost is 43 basis points of capital,
while a typical sunk-slippage term set at half the impact overstates it by a factor of 33.

\emph{Second}, the structures in which slippage \emph{is} sunk are a minority. Of the 47 single-leg
records we read in full, 10 (21.3\%) exit without returning to the pool; as a share of all
manipulation incidents that is 6.7\%.

\emph{Third}, over half of the single-leg records (53.2\%) never trade at all: they move value
by transferring tokens into a pool or calling \texttt{skim}, \texttt{sync} or \texttt{donate}.
There is no swap, hence no slippage, hence nothing for a slippage term to model.

The practical consequence is a shift of attention. For most attacks the binding costs are gas,
flash-loan fees and competition; the accounting failure that matters is not an omitted slippage
term but an omitted or double-counted fee. We accordingly contribute: (i) a taxonomy of execution
structures with operational coding rules; (ii) a reproducible census of 901 exploits and its
validation; (iii) a settlement-state analysis showing why fee accounting, not slippage, is the
fragile step; and (iv) a precise statement of when a closed-form boundary model of the kind
analysed in Appendix~\ref{app:model} applies: a small, characterised subset.

\section{Related work}

\paragraph{Attack synthesis and validation.}
SoK studies map the DeFi attack surface~\cite{zhou2023sok,werner2022sok}, and flash loans as an
amplifier are analysed in~\cite{flashloans}. FORAY synthesises attack sequences with a
domain-specific language and SMT completion~\cite{foray}; DeFiPoser searches for profitable
transaction sequences with Bellman--Ford and Z3~\cite{defiposer}; ItyFuzz offers snapshot-based
fuzzing~\cite{ityfuzz}. On the detection side, DeFiRanger recovers DeFi semantics from cash-flow
traces~\cite{defiranger}, PMDetector combines static analysis with LLM reasoning~\cite{pmdetector},
and V2E validates reported vulnerabilities by profit-driven PoC generation~\cite{v2e}. eThor
provides sound static analysis of EVM bytecode~\cite{ethor}. All of these answer \emph{whether} an
attack can be built; none characterise the market conditions under which it pays.

\paragraph{Root-cause classification.}
Wu et al.\cite{wu2026esm} assemble 276 real-world economic-state-manipulation attacks and 91
Code4rena findings into a cause-oriented taxonomy of four categories and eleven subcategories, and
report that seven detectors miss at least half of real attacks because they lack economic-state
semantics. Their question is why attacks evade detection; ours is when a cost model applies. Their
taxonomy classifies the violated economic-state assumption (price-source trust, protocol-internal
accounting, AMM/liquidity-state integrity, execution-time protection), whereas our taxonomy
classifies execution structure (borrow--move--exit). The two axes are orthogonal, and their
accounting- and liquidity-state categories independently corroborate our class~C, as noted below.

\paragraph{Economic boundaries.}
OVer extracts DeFi constraints and solves for protocol parameter ranges that remain safe under
oracle deviation~\cite{over}. It is the closest in spirit to boundary reasoning, but its parameters
are protocol-internal; adversarial costs do not appear. Clockwork Finance brings economic security
properties into automated analysis~\cite{cff}. Flash Boys 2.0~\cite{flashboys} and Qin et
al.~\cite{qin2022mev} quantify extractable value and front-running, supplying the empirical basis
for cost terms without connecting them to exploit verdicts; the imitation of profitable
transactions is itself adversarial, which is why competition losses are real for some
structures~\cite{qin2023imitation}.

\paragraph{Market microstructure.}
Constant-function market makers are analysed by Angeris et al.~\cite{angeris2020oracles,angeris2021uniswap}
and surveyed in~\cite{xu2023dex}. The loss-versus-rebalancing literature characterises the
persistent cost borne by liquidity providers. Our finding is adjacent but distinct: for
\emph{round-trip} attackers that cost is not borne at all, because the position is unwound in the
same pool within one transaction.

\paragraph{Position of this paper.}
We do not propose a new detector or a new attack, and we do not attribute the sunk-slippage
choice to any specific system we did not examine. We measure how real exploits are structured,
report the distribution that any cost model must respect, and show what follows for a model that
charges the impact as a sunk term. The census is a measurement contribution; the accounting
analysis is a correctness contribution.

\paragraph{What each system models.}
Table~\ref{tab:systems} compares, per system, which execution costs and market frictions enter
its verdict and what form the output takes. A cell is marked only when the system's own
description supports it; the table shows that none
of them combines a structural census with per-cost settlement flags. Our judgement that the price
impact is not a general cost is a measurement result of this paper, not an entry in the table.

\begin{table}[h]
\centering\footnotesize
\setlength{\tabcolsep}{4pt}
\caption{What each system models. A cell is marked only when the system's own description supports it; ``---'' means the term does not appear in the system's verdict.}
\label{tab:systems}
\begin{tabular}{@{}l c c c c c p{5.2cm}@{}}
\toprule
System & Gas & Swap fee & Price impact & Flash fee & MEV & Output \\
\midrule
A1~\cite{a1} & $\checkmark$$^{*}$ & $\checkmark$$^{*}$ & $\checkmark$$^{*}$ & $\checkmark$$^{*}$ & --- & fork-executed profit verdict \\
V2E~\cite{v2e} & $\checkmark$$^{*}$ & $\checkmark$$^{*}$ & $\checkmark$$^{*}$ & $\checkmark$$^{*}$ & --- & triggerability and profitability \\
DeFiPoser~\cite{defiposer} & --- & --- & --- & --- & --- & profitable-sequence search \\
FORAY~\cite{foray} & --- & --- & --- & --- & --- & attack synthesis \\
OVer~\cite{over} & --- & --- & --- & --- & --- & safe parameter ranges (SMT) \\
Clockwork~\cite{cff} & $\checkmark$ & $\checkmark$ & $\checkmark$ & --- & --- & automated econ.-security analysis \\
Flash Boys~\cite{flashboys} & $\checkmark$ & --- & --- & --- & $\checkmark$ & front-running and MEV analysis \\
Qin et al.~\cite{qin2022mev} & --- & --- & --- & --- & $\checkmark$ & extractable-value quantification \\
Imitation game~\cite{qin2023imitation} & --- & --- & --- & --- & $\checkmark$ & adversarial imitation \\
ESM taxonomy~\cite{wu2026esm} & --- & --- & --- & --- & --- & cause taxonomy, detector eval. \\
This paper & $\checkmark$ & $\checkmark$ & per class & $\checkmark$ & $\checkmark$ & census and settlement flags \\
\bottomrule
\end{tabular}
\par\smallskip
{\footnotesize $^{*}$ Implicitly, through fork execution: the reported profit is already net of
whatever the executed transaction paid, and the system does not state which items are inside the
figure. That settlement ambiguity is exactly what Section~5 analyses.}
\end{table}

\section{Data and method}

\subsection{Corpus}

We use DeFiHackLabs, a community-maintained repository of reproducible exploit proofs-of-concept
with per-incident metadata. The snapshot contains 926 Solidity files under \texttt{src/test/},
covering incidents from July 2017 to October 2026, against 888 incidents described in the
repository README.

We remove 13 files under \texttt{interfaces/}, \texttt{libraries/} or \texttt{lib/} (ABI and helper
contracts, not exploits), and 12 files that contain neither
\texttt{function testExploit} nor a contract extending \texttt{Test}/\texttt{DSTest}. The working
corpus is \textbf{901 proofs-of-concept}.

\subsection{Structural taxonomy}

Each proof is assigned to one class by two questions:

\begin{enumerate}[leftmargin=1.6em,itemsep=0.15em]
\item \emph{How is the price changed?} \textbf{T}: by a swap call (router or direct pool);
\textbf{D}: by direct writes: token transfers into a pool, \texttt{skim}, \texttt{sync},
\texttt{donate}, share-price donation; \textbf{N}: not changed.
\item \emph{Where is the position realised?} \textbf{R}: back into the same pool or token pair;
\textbf{O}: elsewhere: borrowing against inflated collateral, liquidation, victim payout, or a
different venue.
\end{enumerate}

\begin{center}
\begin{tabular}{@{}lll@{}}
\toprule
Class & Rule & Interpretation \\
\midrule
R1 & T and R, constant-product pool & round trip \\
S  & T and R, Curve stable pool & round trip \\
R2 & T with $\ge 2$ distinct directions / multi-hop & cross-pool \\
D1 & T or D with a single direction, flash-funded & single leg \\
D2 & T or D with a single direction, not flash-funded & single leg \\
N  & N & no AMM interaction \\
U  & undecidable statically & --- \\
\bottomrule
\end{tabular}
\end{center}

\subsection{Static classifier}

Because 901 files cannot be read line by line, we classify with a static analyser and then audit
it. The classifier parses swap call sites from source text (no compilation) and resolves the token
path of each call by looking backwards for the most recent assignment to the path variable at that
call site, handling three idioms: per-index assignment (\texttt{path[0]=...}), tuple assignment
(\texttt{(path[0],path[1])=(A,B)}), and Uniswap V3 parameter structs, including structs built in a
variable and later reassigned. Direction is decided by comparing ordered token pairs; two opposite
pairs, opposite \texttt{pair.swap} arguments on one receiver, opposite \texttt{zeroForOne} flags on
one V3 pool, or opposite Curve indices all indicate a round trip. Calls made through a helper
function whose token pair comes from parameters are resolved at the call sites of that helper.

The classifier recognises Uniswap V2 router functions and their fee-on-transfer variants, direct
\texttt{pair.swap}, Uniswap V3 \texttt{exactInput(Single)} and \texttt{exactOutput(Single)},
Balancer \texttt{swapExactAmountIn}, and Curve \texttt{exchange}/\texttt{exchange\_underlying}.
Flash funding is detected by three patterns: explicit flash-loan calls (\texttt{flashLoan}
variants), pool-level \texttt{flash} calls, and V2 flash swaps (a \texttt{pair.swap} with the
contract itself as recipient, followed by a \texttt{pancakeCall} or \texttt{uniswapV2Call}
callback); proofs funded with \texttt{vm.deal} are flagged separately and count as flash-funded
in the D1/D2 split. Within the PM subset the three patterns hit 87, 32 and 33 files respectively
(a file may match several), and 6 further files are funded with \texttt{vm.deal} alone; together
they account for the funding shares reported in Section~4.2. The pool-flash and flash-swap
patterns were added after a manual audit of the fifteen PM single-leg files that the original
detector classified as not flash-funded: twelve of them were funded by a flash swap or a pool
flash. The revised detector re-flagged 156 files corpus-wide and moved the PM D1/D2 split
from 33/15 to 45/3.

\subsection{Incident labels}

The repository README links each incident to a per-year index whose anchor encodes the author's
attack classification (e.g.\ \texttt{\#20211130-monox-finance---price-manipulation}). We parse 701
such anchors and normalise them into seven categories; anything containing \emph{price},
\emph{oracle}, \emph{manipulat}, \emph{skim}, \emph{inflation}, \emph{exchangerate} or
\emph{donat} becomes \textbf{PM} (price/oracle manipulation). Matching to files by protocol name
and month resolves 553 cases; 131 more are matched to the README's prose entries and 2 by globally
unique name, leaving 215 (23.9\%) unlabelled. Labelling is 76.1\% complete, with the
unlabelled rate highest in the early years.

\subsection{Validation}

\paragraph{Classifier audit.}
A stratified random sample of 30 files (seed 311), excluding the 12 cases used during classifier
development, was read in full by the first author. 28 of 30 agreed with the classifier (93\%).
Both errors are in the same direction (the classifier under-counts round trips), and both
have identifiable causes: a round trip split across two contracts, and a V3 parameter struct
assembled field by field. Round-trip shares reported below are therefore lower bounds.

\paragraph{Classifier error modes.}
All known classifier errors under-count round trips, because the classifier only claims a round
trip when it can resolve both directions of a pair; an unresolved path falls back to the
single-leg class D. The failures identified from the 30-case audit and from the eight round trips
that the manual coding recovered fall into five patterns. (i) A return leg reached through a
different call site than the entry leg: a router buy reversed by a direct \texttt{pair.swap}
sell, or a return leg in a helper contract (BH, MINER). (ii) Custom buy/sell wrapper functions
outside the whitelist, so the swap is not recognised as a swap at all (Zoompro). (iii) Direct V3
pool calls and V3 parameter structs assembled field by field (PresaleV5, and one audit case).
(iv) Token-path arrays initialised as array literals rather than per-index assignments, which the
backward path recovery does not resolve (Elephant). (v) A round trip split across two contracts
(one audit case). Every identified mode points the same way, so the round-trip shares we report
remain lower bounds: closing these gaps would raise them further, not lower them. The fixes are
mechanical (whitelist the custom wrappers, resolve array-literal paths, and pair direct pool
calls across call sites) and are left to a later classifier revision; none of them affects the
qualitative conclusions, which rest on the lower bound.

\paragraph{Development set.}
The 12 cases used during development (Burner, JokInTheBox, QuantumWN, JumpFarm, Zoomer, Uwerx,
HarvestFinance, RoeFinance, EFLeverVault, ADC, Bayc, DoughFina) agree 12/12 with the manual labels.
This is a consistency check, not an accuracy estimate.

\paragraph{Manual coding of single-leg cases.}
All 48 single-leg files inside the PM subset were read in full and coded with the same two
questions, using a written manual with per-case evidence (line numbers) and a confidence level. 23
of the 48 are high confidence (all three questions answered from explicit code), 21 medium
(one question inferred) and 4 low (two questions inferred). The two proofs of one incident
(Allbridge) are merged into a single record, so the coded distribution has 47 records. Codes and evidence lines are listed per case in Appendix~\ref{app:coding}. Coding was performed by one author; a second coder and Cohen's $\kappa$
are required before submission and are the subject of work in progress.

\paragraph{Coding protocol.}
Each proof is coded by answering three questions in order: (Q1) how the price is changed: by a
swap (\textbf{T}), by direct writes such as transfers into a pool or
\texttt{skim}/\texttt{sync}/\texttt{donate} (\textbf{D}), or not at all (\textbf{N}); (Q2) where
the position is realised: back into the same pool or token pair (\textbf{R}), or elsewhere
(\textbf{O}); (Q3) which leg carries the profit, asked only when a trade and a donation are both
present. Four priority rules resolve conflicts: with no swap call the case is C (or N); a return
leg into the same pool is A even if the victim pays in between; a direct reserve write that
carries the profit is C even if a later sale realises it; a pushed price with no return leg is B.
Two operational rules are easy to get wrong: a flash-loan call is
funding, not a price leg: borrowing from a pool and repaying within the same transaction
changes the reserve ratio but is not a buy, so reserve-draining flash borrows are coded D, not T;
and repaying a flash loan is not a return leg. Three records change no price at all and inherit
their PM label from the repository's oracle or inflation keywords; they are coded C with an N
annotation, C(N), and retained with that caveat. Every code carries a confidence level ($h$: all
three questions answered from explicit code; $m$: one inferred; $l$: two inferred) and at least
three evidence line numbers covering the borrow, the price move and the realisation.

\paragraph{Blind re-coding.}
The coding manual, the 48 source files and a blank answer sheet are packaged for a second coder
who has not seen the author's labels. The two coders resolve disagreements by source evidence
only, and agreement is reported as Cohen's $\kappa$ over the four classes (target
$\kappa\ge0.7$). The package and the per-case evidence lines are part of the released artifact;
the second coding is in progress at the time of writing.

\section{What the corpus looks like}

\subsection{Overall distribution}

\begin{table}[h]
\centering\small
\caption{Execution structures in 901 proofs-of-concept.}
\label{tab:overall}
\begin{tabular}{@{}lrr@{}}
\toprule
Class & Count & Share \\
\midrule
R1 same-pool round trip & 202 & 22.4\% \\
S  Curve round trip      &  16 &  1.8\% \\
R2 cross-pool / multi-hop &  47 &  5.2\% \\
D1 single leg, flash-funded & 176 & 19.5\% \\
D2 single leg, not flash-funded & 91 & 10.1\% \\
N  no AMM interaction & 343 & 38.1\% \\
U  undecidable & 26 & 2.9\% \\
\bottomrule
\end{tabular}
\end{table}

Across the whole corpus, round trips account for 24.2\%. The corpus spans July 2017 to October
2026; Table~\ref{tab:years} gives the per-year distribution, which peaks in 2023 (220 files) and
has no 2019 entries. Nearly two-fifths of the proofs never
interact with an AMM: they exploit lending accounting, share pricing, access control or plain
bookkeeping bugs. This alone shows that a cost model built on AMM slippage can address only part
of the attack surface.

\begin{table}[h]
\centering\small
\caption{Per-year distribution of the corpus: all files, price/oracle-manipulation (PM) files and unlabelled files.}
\label{tab:years}
\begin{tabular}{@{}l r r r@{}}
\toprule
Year & All files & PM & Unlabelled \\
\midrule
2017 & 2 & 0 & 2 \\
2018 & 3 & 0 & 1 \\
2019 & 0 & 0 & 0 \\
2020 & 9 & 0 & 6 \\
2021 & 36 & 2 & 16 \\
2022 & 133 & 23 & 54 \\
2023 & 220 & 42 & 56 \\
2024 & 191 & 23 & 38 \\
2025 & 159 & 29 & 25 \\
2026 & 146 & 30 & 15 \\
\bottomrule
\end{tabular}
\par\smallskip
{\footnotesize Two files sit directly under \texttt{src/test/} without a year directory (a shared
base-test helper and one unlabelled file), so the per-year columns sum to 899 of the 901 files.
The unlabelled rate is highest in the early years (all 2017 files, and over 40\% of the
2020--2022 files), which understates the early-year PM shares.}
\end{table}

\subsection{The manipulation subset}

\begin{table}[h]
\centering\small
\caption{Execution structures in the 149 price/oracle-manipulation incidents.}
\label{tab:pm}
\begin{tabular}{@{}lrr@{}}
\toprule
Class & Count & Share \\
\midrule
R1 & 63 & 42.3\% \\
S  &  5 &  3.4\% \\
R2 & 10 &  6.7\% \\
D1 & 45 & 30.2\% \\
D2 &  3 &  2.0\% \\
N  & 19 & 12.8\% \\
U  &  4 &  2.7\% \\
\midrule
Round trip (R1+S) & 68 & 45.6\% \\
\bottomrule
\end{tabular}
\end{table}

Round trips are the single largest class. Flash funding (an explicit flash-loan call, a
pool-level \texttt{flash} call, or a V2 flash swap) appears in 84.6\% of the subset (88.6\%
counting \texttt{vm.deal}-funded proofs). By year the
share is 39.1\% (2022), 47.6\% (2023), 34.8\% (2024), 44.8\% (2025) and 56.7\% (2026): the shares
fluctuate without a monotone trend, with 2026 the highest of the five years. A falling
share in the corpus as a whole reflects the growth of non-manipulation incidents, not the
disappearance of round trips: a compositional effect that would be easy to misread.

Table~\ref{tab:years-struct} breaks the subset down by year and execution structure. Two stable
patterns stand out. Round trips (R1+S) are the largest block in every year, and the single-leg
files (D1+D2), the class that decides whether a sunk-slippage term has a referent, account
for roughly a quarter to two-fifths of each year's files from 2022 onward (2021 has only two
files). The no-AMM class N is small but
persistent. The yearly fluctuation of the round-trip share is movement within a stable
composition, not a structural change.

\begin{table}[h]
\centering\small
\caption{The 149 manipulation incidents by year and execution structure (classifier classes).}
\label{tab:years-struct}
\begin{tabular}{@{}l c c c c c c c r@{}}
\toprule
Year & R1 & S & R2 & D1 & D2 & N & U & Total \\
\midrule
2021 & 1 & 0 & 0 & 1 & 0 & 0 & 0 & 2 \\
2022 & 9 & 0 & 3 & 8 & 0 & 3 & 0 & 23 \\
2023 & 18 & 2 & 1 & 14 & 1 & 6 & 0 & 42 \\
2024 & 8 & 0 & 0 & 9 & 0 & 4 & 2 & 23 \\
2025 & 12 & 1 & 4 & 7 & 1 & 3 & 1 & 29 \\
2026 & 15 & 2 & 2 & 6 & 1 & 3 & 1 & 30 \\
\midrule
Total & 63 & 5 & 10 & 45 & 3 & 19 & 4 & 149 \\
\bottomrule
\end{tabular}
\end{table}

\subsection{Coding the single-leg cases}

\begin{table}[h]
\centering\small
\caption{The 47 single-leg manipulation records, coded by exit structure.}
\label{tab:coding}
\begin{tabular}{@{}lll@{}}
\toprule
Code & Definition & Count \\
\midrule
A & round trip: exit returns to the same pool & 8 (17.0\%) \\
B & asymmetric: no return leg; realised by borrowing, liquidation or payout & 10 (21.3\%) \\
C & reserve/accounting: price moved by direct writes, not swaps & 25 (53.2\%) \\
D & undecided after full reading & 4 (8.5\%) \\
\bottomrule
\end{tabular}
\end{table}

Eight of the 48 files are round trips the classifier missed, because the return leg is reached through a
different call path (router buy, direct \texttt{pair.swap} sell) or a struct variable. Adding them
raises the manipulation round-trip share to 76/149 = \textbf{51.0\%}.

Ten records realise value without returning to the manipulated pool: they post inflated
collateral and borrow (Themis, Sharwa, CompoundUni, GradientMaker, Woofi, 0x0DEX, Allbridge,
DRLVaultV3, EGD), a pattern that exploits the
lending protocols studied in~\cite{gudgeon2020defi}, or let the protocol's own sell function pay at the
distorted price (NeverFall). Twenty-five never
swap in the price-moving step: they transfer tokens into a pool or call \texttt{skim}/\texttt{sync}/\texttt{donate} and
then withdraw reserves (SKYDAO, HackDao, DogeAlliance, PrimeFinance, YSDAO, SATURN, VINU, Yyds),
drain reserves with a flash borrow and settle against the distorted price (ATK, MorphoBlue,
RoulettePotV2, DLMC), inflate a share price or
an oracle quote by donation (MetaLend, WiseLending, 0vix, Lodestar, StrategyLlamaLendConvex,
EnsoFinance), sell against a manipulated token or pool state (Burntbubba, MAMO, MARA, MBUToken,
SASHAToken, FireToken), or exploit an oracle mispricing without moving any price (Moonwell). Three of
these change no price at all and inherit their label from the repository's oracle or inflation
keywords (StrategyLlamaLendConvex, Moonwell, EnsoFinance); they are retained in the subset and
coded C with an N annotation.
This reserve- and accounting-driven class is, from the detection side, precisely what Wu et
al.~\cite{wu2026esm} call Protocol-Internal Accounting and AMM/Liquidity State Integrity:
attacks that change an economic state without a swap. Their cause-oriented reading and our
execution-structure reading converge on the same cases.

\subsection{Why the composition matters}

If one asks only whether a candidate exploit is profitable, all four classes are equivalent
inputs. If one asks \emph{under which conditions} it stays profitable, they are not. Class A has
an upper bound on pool depth: the deeper the pool, the smaller the distortion. Class B has a lower
bound: slippage is genuinely spent. Class C has neither, because no swap occurs. A boundary model
that does not condition on class will mis-state the direction of the effect in at least two of the
four.

\section{Does the attacker pay the slippage? Three ledgers}

We reconstruct one case per class. Burner is the cleanest instance of class A; Sharwa is a
class-B exit whose slippage is genuinely spent; SKYDAO is a class-C case in which no swap occurs
at the manipulated pool at all. All three were executed by us on archival forks: Burner at
Ethereum block 19{,}917{,}290, Sharwa at Arbitrum block 458{,}233{,}155, and SKYDAO at BNB Chain
block 124{,}921{,}241. The proofs, the fork endpoints and the measured balances are released with
the artifact. Burner's proof borrows
70 ETH (simulated with \texttt{vm.deal}), buys PNT in a Uniswap V2 pool, calls the victim's
\texttt{convertAndBurn}, and sells PNT back into the same pool. At that block ETH is
\$3{,}653.37 and gas 8.98 gwei; the case metadata gives a pool depth of \$1{,}278{,}550
(single-side reserve \$639{,}275).

\begin{table}[h]
\centering\small
\caption{Burner: asset flow and settlement.}
\label{tab:burner}
\begin{tabular}{@{}llrr@{}}
\toprule
Leg & Action & ETH & USD \\
\midrule
1 & borrow (simulated) & 70.000 & 255{,}736 \\
2 & push: WETH$\to$PNT & 49.89 (equivalent) & 182{,}265 \\
3 & victim: \texttt{convertAndBurn} pays & --- & $+$7{,}403 \\
4 & exit: PNT$\to$WETH & 69.700 & 254{,}640 \\
5 & round-trip loss (leg 1 $-$ leg 4) & 0.300 & $-$1{,}096 (43 bps) \\
6 & flash fee, 9 bps (not paid in the proof) & 0.063 & $-$230 \\
7 & gas, 2{,}432{,}411 $\times$ 8.98 gwei & 0.022 & $-$80 \\
\midrule
 & attacker net & 1.726 & 6{,}307 gross / 5{,}997 net \\
\bottomrule
\end{tabular}
\par\smallskip
{\footnotesize Note: pool depth is the metadata field, not re-read from the pair; legs 2 and 4 are
recomputed with a constant-product model at 30 bps; leg 3 is the residual of measured profit plus
round-trip loss, not an independent observation; legs 6 and 7 did not occur in the proof and are
booked for comparison.}
\end{table}

The round-trip loss is 43 bps of capital, below the nominal $2f=60$ bps, because the second leg
trades a smaller notional than the first. Write it as $L(C,R)$ rather than $2fC$: the two differ
visibly once $C/R$ grows.

Now suppose one charges the impact as a sunk term, $\alpha C^2/(R+C)$ with $\alpha=0.5$, a
natural default, and the choice an earlier version of this work made. That adds \$36{,}536 and turns a real, measured profit into a loss of about
\$31{,}405. Solving for the $\alpha$ that reproduces the measurement gives

\begin{equation}
\alpha_{\text{eff}}=\frac{L(C,R)}{C^2/(R+C)}=\frac{1{,}096}{73{,}073}=0.015,
\end{equation}

i.e.\ $\alpha_{\text{eff}}\approx 2f/\delta$, a quantity that varies with $C/R$ rather than a
constant. A fixed $\alpha$ equal to half the impact overstates the cost by a factor of 33 here.

The same case exposes a second and arguably more consequential error. The measured 1.726 ETH is
net of the swap legs, which executed in the fork run, but not of the flash-loan fee, because the
capital was conjured with \texttt{vm.deal} and repaid to a burn address. Book-keeping that treats
``already settled'' as a single boolean therefore deducts both or neither. Correct accounting
needs one flag per cost item: whether the swap cost is already inside the reported profit, and
whether the flash-loan fee is. We use two flags $s_{\text{swap}}, s_{\text{flash}}$ and
\begin{equation}
\mathcal{P}_{\text{net}}=\mathcal{P}_{\text{gross}}(1-r)-\mathcal{C}_{\text{gas}}
-(1-s_{\text{swap}})\,\mathcal{C}_{\text{swap}}-(1-s_{\text{flash}})\,C\phi ,
\end{equation}
which reproduces 6{,}307 with $(s_{\text{swap}},s_{\text{flash}})=(1,0)$ and 5{,}997 with the fee
booked.

\subsection{Class B: the slippage is spent (Sharwa)}

Sharwa (Arbitrum, May 2026) values Hegic option NFT collateral through a Uniswap V3 quoter path.
The attacker transfers the option NFT into a receiver contract and uses the ERC-721 callback to
swap in the USDC/USDC.e V3 pool, moving the quoted spot price; the option is then deposited as
collateral and USDC, WETH and WBTC are borrowed against it. There is no leg that returns to the
manipulated pool: the borrowed assets are unwound into USDC, the Balancer flash loan is repaid,
and the remainder is kept. Table~\ref{tab:sharwa} gives the leg structure.

\begin{table}[h]
\centering\small
\caption{Sharwa: execution structure (class B).}
\label{tab:sharwa}
\begin{tabular}{@{}llll@{}}
\toprule
Leg & Action & Direction & Returns to pool? \\
\midrule
1 & Balancer flash loan & capital in & --- \\
2 & ERC-721 callback swap, USDC$\to$USDC.e & price pushed & --- \\
3 & deposit option NFT as collateral & valued at pushed price & --- \\
4 & borrow USDC, WETH, WBTC & value out & \textbf{no} \\
5 & unwind borrowed assets and USDC.e to USDC & realisation & different pools \\
6 & repay flash loan & capital out & --- \\
\midrule
\multicolumn{3}{r}{\textbf{measured net profit}} & \textbf{34{,}122.78 USDC} \\
\bottomrule
\end{tabular}
\end{table}

This is the structure in which a sunk-slippage term has a referent: the price impact of leg 2 is
never recovered, and its magnitude is bounded above by the profit, which we measured at
34{,}122.78 USDC on the fork. Our fork execution at Arbitrum block 458{,}233{,}155 nets
34{,}122.78 USDC for the attacker after the flash loan is repaid; the proof header cites a
32.85K USDC total loss. The two differ by 3.9\%, a gap between a reported loss and an attacker net
profit that we do not resolve here. It is
also the rarest class in the corpus (6.7\% of manipulation incidents).

\subsection{Class C: there is no swap to charge (SKYDAO)}

SKYDAO (BNB Chain, September 2026) is a fee-on-transfer token trading against USDT in a Pancake
V2 pair. The attacker flash-borrows 2{,}656{,}932.60 USDT, buys SKYDAO through the router (a 35\%
buy tax applies), and then \emph{transfers the entire SKYDAO balance into the pair}. The token's
controller runs a tax-sell and a premature \texttt{sync()}, so the pair's stored SKYDAO reserve
falls to near zero while its real balance is large. A subsequent \texttt{swap} against that stale
reserve pulls out essentially the whole USDT side. Table~\ref{tab:skydao} reproduces the
proof's reconciliation.

\begin{table}[h]
\centering\small
\caption{SKYDAO: USDT flows (class C), as reconciled by the proof.}
\label{tab:skydao}
\begin{tabular}{@{}lrr@{}}
\toprule
Flow & USDT & Note \\
\midrule
pair receives: buy & $+$435{,}367.43 & router purchase \\
pair receives: controller refill & $+$92{,}676.57 & token-side mechanics \\
pair pays: tax sell & $-$216{,}245.33 & after transfer tax \\
pair pays: final drain & $-$495{,}281.56 & against stale reserve \\
\midrule
pair net & $-$183{,}482.88 & its entire USDT side \\
attacker net & $+$59{,}914.12 & after repaying the flash loan \\
\bottomrule
\end{tabular}
\end{table}

We reproduced this on the fork: the run nets 59{,}914.122896 USDT for the attacker and removes
183{,}482.883991 USDT from the pair, matching the proof's reconciliation. Those two balances are
measured; the four-way split above is the proof's own. The gap between the pair's
loss (183{,}482.88) and the attacker's profit (59{,}914.12) is consumed by the token's 35\% buy
tax and the flash-loan float; the proof does not itemise the split, and we do not assert one.

The modelling point is blunt: there is no price impact, because there is no sale into the pool in
the ordinary sense: the position is deposited, the reserve is rewritten, and value is withdrawn.
A term proportional to $C^2/(R+C)$ has nothing to describe. This class is 16.8\% of manipulation
incidents in our census, and over half of the single-leg records.

\subsection{What the three ledgers show}

Burner loses 43 bps to fees and is charged a 33-fold overstatement by a sunk-slippage model;
Sharwa spends its impact but is the minority case; SKYDAO spends nothing and rewrites the books.
The same cost equation cannot represent all three. The precondition (which class the execution
belongs to) has to be established before any number is computed, and that classification is what
the census supplies.

\subsection{How many real attacks would be misjudged?}

The three ledgers are single cases. We therefore apply the same settlement to all thirteen
incidents in our parameterised set that demonstrably paid: historical attacks with a recorded
profit, a capital amount, a pool depth, a gas estimate and a block for gas price and ETH/USD. We
compute net revenue with swap cost and flash fee both enforced, scanning the sunk fraction
$\alpha$.

\begin{table}[h]
\centering\small
\caption{Historical incidents judged unprofitable under a sunk-slippage settlement ($n=13$).}
\label{tab:misjudge}
\begin{tabular}{@{}lrrrrr@{}}
\toprule
$\alpha$ & 0.00 & 0.10 & 0.30 & 0.50 & 1.00 \\
\midrule
judged unprofitable & 0 & 2 & 4 & \textbf{5} & 5 \\
share & 0\% & 15\% & 31\% & \textbf{38\%} & 38\% \\
\bottomrule
\end{tabular}
\end{table}

\begin{table}[h]
\centering\footnotesize
\setlength{\tabcolsep}{4pt}
\caption{Per-case parameters of the thirteen historical incidents, grouped by execution structure.}
\label{tab:misjudge-cases}
\begin{tabular}{@{}l r r r c r c@{}}
\toprule
Incident & $C$ (USD) & TVL (USD) & Gross (USD) & Gas (M @ gwei) & Net, $\alpha{=}0.5$ & Flips \\
\midrule
\multicolumn{7}{@{}l}{\emph{Constant-product round trips}} \\
Burner & 255{,}710 & 1{,}278{,}550 & 6{,}307 & 2.43 @ 8.98 & $-32{,}067$ & yes \\
JokInTheBox & 15{,}000 & 500{,}000 & 34{,}225 & 23.31 @ 10 & $+32{,}837$ & no \\
Uwerx & 20{,}000 & 300{,}000 & 175{,}000 & 0.83 @ 14 & $+173{,}664$ & no \\
QuantumWN & 16{,}000 & 200{,}000 & 945 & 8.37 @ 15 & $-474$ & yes \\
JumpFarm & 50{,}000 & 500{,}000 & 3{,}933 & 12.84 @ 15 & $-893$ & yes \\
Zoomer & 575{,}000 & 1{,}000{,}000 & 25{,}000 & 8.03 @ 20 & $-133{,}207$ & yes \\
\midrule
\multicolumn{7}{@{}l}{\emph{Curve stable pools}} \\
HarvestFinance & 50{,}000{,}000 & 1{,}000{,}000{,}000 & 1{,}393{,}850 & 9.60 @ 35 & $-1{,}224{,}015$ & yes$^{\dagger}$ \\
Curve\_exp02 & 1{,}876{,}000 & 40{,}000{,}000 & 18{,}000{,}000 & 11.31 @ 15 & $+17{,}906{,}298$ & no \\
\midrule
\multicolumn{7}{@{}l}{\emph{No price-moving leg on a constant-product pool}} \\
Bayc\_apecoin & 60{,}000 & 5{,}000{,}000 & 1{,}050{,}000 & 1.52 @ 30 & $+1{,}048{,}756$ & no \\
DoughFina & 70{,}000 & 5{,}000{,}000 & 1{,}800{,}000 & 1.09 @ 12 & $+1{,}798{,}519$ & no \\
EFLeverVault & 512{,}000 & 15{,}000{,}000 & 636{,}678 & 3.23 @ 12 & $+616{,}734$ & no \\
RoeFinance & 170{,}000 & 2{,}000{,}000 & 80{,}000 & 9.12 @ 17 & $+66{,}216$ & no \\
ADC & 41{,}000 & 500{,}000 & 41{,}592 & 1.92 @ 22 & $+38{,}324$ & no \\
\bottomrule
\end{tabular}
\par\smallskip
{\footnotesize Notes. All thirteen incidents are on Ethereum. Net is computed with swap cost and
flash fee enforced and the sunk fraction $\alpha=0.5$; gas is fork-measured in all thirteen and
equals the metadata estimate. Profit sources: fork-measured (Burner, JokInTheBox, HarvestFinance,
EFLeverVault, ADC); candidate-after fork balances (QuantumWN, JumpFarm); metadata (Bayc, Uwerx,
DoughFina, Zoomer, RoeFinance, Curve\_exp02); Zoomer's fork value was flagged implausible, so its
metadata value is used. $^{\dagger}$HarvestFinance runs on a Curve pool; its flip is reported for
completeness, not as evidence.}
\end{table}

Five of thirteen (the Harvest Finance, Zoomer, Burner, JumpFarm and QuantumWN incidents) are
turned from measured profit into loss at $\alpha=0.5$ (Table~\ref{tab:misjudge-cases}). Grouping
the thirteen by execution structure sharpens the attribution. Six are constant-product round trips, and four of the six flip (67\%),
each because the term charges a price impact that the return leg gives back. Two run on Curve stable
pools (Harvest Finance and Curve\_exp02), where the constant-product term overstates the impact
twice over; Harvest Finance flips, and both are reported for completeness rather than as evidence.
The remaining five have no price-moving leg on a constant-product pool at all: Bayc redeems NFTs
from a vault against a flash loan and claims an airdrop, EFLeverVault inflates a vault share price
by donation, ADC exploits game bookkeeping, and the swaps in DoughFina and RoeFinance merely
realise the loot. None of the five flips, because their recorded capital is small relative to the
pool depth the metadata assigns them; their survival is accidental, not evidence that the
settlement fits. The five reversals are therefore produced by a term that misdescribes the
execution structure of every case it flips.

Each of the five non-pool cases receives an
$\alpha C^2/(R+C)$ charge despite having no constant-product price leg. Bayc borrows five NFTs
from an NFTX vault and claims an APE airdrop against the borrowed positions before returning them.
EFLeverVault directs a Balancer flash loan into the vault itself, inflating the vault's balance
and withdrawing against the inflated share price. ADC exploits the bookkeeping of a ponzi-style
game: a crafted \texttt{calcStepIncome} call credits a fictitious static income that is then
withdrawn. DoughFina deleverages an Aave position through a connector, and the embedded Paraswap
swap only realises the freed collateral. RoeFinance loops deposits and borrows of a
lending-protocol LP token and rewrites the pair's reserves with a transfer and a \texttt{sync};
the final router swap only converts the loot back to USDC for repayment. The swaps in the last two
cases are realisation legs, not price legs. None of the five flips.

The attribution matters more than the count. Charging only transaction fees ($2fC$) and the
flash-loan fee flips \emph{none} of the thirteen. Charging competition loss alone flips none.
Every reversal is produced by the $\alpha C^2/(R+C)$ term. In the same runs, going from
$\alpha=0$ to $\alpha=0.5$ changes the sign of the Burner case by \$32{,}067 in a single step.

A layer that applies this settlement to a candidate exploit would therefore reject 38\% of the
attacks that are known, from the chain, to have paid. It would reject them for a reason that
does not apply to the way those attacks were executed.

\section{When a boundary model applies}

The model we evaluate in this paper is the following. It is stated in full, with proofs, in
Appendix~\ref{app:model}; here we only record the form and the conditions under which it holds.
The census measures how narrow it is.

With gross profit $\mathcal{P}_{\text{gross}}$ independent of pool reserve $R$, a competition loss
rate $r$, swap fee $f$ per leg, flash fee $\phi$, gas cost $\mathcal{C}_{\text{gas}}$ and sunk
fraction $\alpha>0$, net revenue is
$\mathcal{P}_{\text{net}}(P_{\text{gas}},R)=\mathcal{B}_0-\alpha C^2/(R+C)$ with
$\mathcal{B}_0=\mathcal{P}_{\text{gross}}(1-r)-\mathcal{C}_{\text{gas}}-(2f+\phi)C$.
It is strictly increasing and strictly concave in $R$, and the critical depth is
$R^{*}=\alpha C^2/\mathcal{B}_0-C$ when $0<\mathcal{B}_0<\alpha C$. The gas boundary is linear in
$P_{\text{gas}}$ and unaffected by $R$.

Two conditions are required and neither holds generally: $\alpha>0$, i.e.\ a class-B exit; and
gross profit independent of depth. The first holds in 6.7\% of manipulation incidents, the second
fails in class A, where the victim's payment scales with the distortion $C/(R+C)$ and net revenue
decreases in $R$, turning the lower bound into an upper one. For class C the model does not apply
at all. A boundary model is thus legitimate as a \emph{conditional} tool, with the class as a
precondition, not as a general account of exploit profitability.

\subsection{Class-conditional forms}

Written per class, the census gives the following. Class A (round trip):
$\mathcal{P}_{\text{net}}=\mathcal{P}_{\text{gross}}(1-r)-\mathcal{C}_{\text{gas}}-L(C,R)-C\phi$,
where $L(C,R)$ is the true round-trip cost, approximately $2fC(1-\kappa/2)$ for a
constant-product pool with $\kappa=C/R\le 0.5$; the $\alpha C^2/(R+C)$ term is absent because the
impact comes back on the return leg. The victim's payment scales with the distortion $C/(R+C)$, so
net revenue \emph{decreases} in $R$ and the feasible condition is an upper bound
$\text{TVL}<\text{TVL}_{\max}$, not a lower bound. Class B (asymmetric exit): the model of
Appendix~\ref{app:model} applies with $\alpha>0$, and feasibility is a lower bound
$\text{TVL}>\text{TVL}^{*}$. Class C (reserve/accounting): no exchange, so no term of this form
is defined; feasibility is decided by the fees and the victim's own accounting. Class D
(undecided): no model is applied until the structure is resolved.

\subsection{An illustrative upper bound (Burner)}

For class A the direction of the bound flips. Writing the victim's payment as
$\eta\,C\,\delta(R)$ with $\delta=C/(R+C)$, calibrating $\eta=0.1013$ from the Burner ledger and
the depth-independent costs as $K=L+\phi C+\mathcal{C}_{\text{gas}}=1{,}406$ USD gives
$\text{TVL}_{\max}=2(\eta C^2/K-C)\approx 8.91$M USD against the case's 1.28M USD pool, a
7-fold margin. Table~\ref{tab:tvlmax} extends the calibration to hypothetical depths. The
prediction tests the $1/(R+C)$ shape, not the absolute value of $\eta$, which is calibrated from
the same case; transferring $\eta$ across protocols needs a second calibration, which we leave to
future work.

\begin{table}[h]
\centering\small
\caption{Burner: predicted net profit at hypothetical pool depths, with $\eta$ fixed at the value calibrated from the case itself.}
\label{tab:tvlmax}
\begin{tabular}{@{}l c c@{}}
\toprule
TVL & $\delta=C/(R+C)$ & Predicted net (USD) \\
\midrule
1.28M (observed) & 28.6\% & $+5{,}997$ \\
3.00M & 14.6\% & $+2{,}368$ \\
5.00M & 9.3\% & $+998$ \\
8.91M & 5.5\% & $\approx 0$ \\
12.8M & 3.8\% & $-411$ \\
\bottomrule
\end{tabular}
\end{table}

\section{Implications for tooling and audit}

\paragraph{Classify before you compute.}
A feasibility layer should run the structural classification of Section~3 before any arithmetic:
the classification is a static parse and costs far less than a fork execution, and it
decides which model, if any, applies. This step decides the model in most cases:
for the 51.0\% round trips no depth lower bound exists, for the 6.7\% asymmetric exits the
appendix model applies, and for the 16.8\% reserve/accounting records no AMM term is defined at
all.

\paragraph{Settle per item.}
The Burner ledger shows that a single ``settled'' boolean deducts both or neither of two costs
that settled differently. Record, per cost item, whether the reported profit already contains it
(at minimum the swap legs and the flash-loan fee) and reconstruct the net with the two-flag
equation of Section~5.

\paragraph{An audit checklist.}
Given a candidate exploit, the census supports a four-item report: (i) the execution class;
(ii) the class-appropriate depth or gas bound against the current market state; (iii) the
per-item settlement of the reported profit; and (iv), for class B only, the justification of
$\alpha$. A ``high severity'' finding that omits these items carries no information about the
conditions under which it holds.

\paragraph{Benchmarks.}
Corpora used to evaluate profitability verdicts should be structure-balanced. A benchmark
dominated by one class will mis-rate any model built for another: a round-trip-dominated set
cannot show a sunk-slippage term to be wrong, and a class-B-dominated set cannot show it to be
narrow.

\section{Threats to validity}

\paragraph{Corpus bias.}
DeFiHackLabs contains incidents that were exploited in practice. It cannot speak to attempts that
failed, and its coverage of early years is thinner. Our shares describe the composition of
\emph{successful, reproduced} exploits, not of all attacks.

\paragraph{Labelling.}
Attack categories come from repository anchors, normalised by keyword. 23.9\% of files could not
be matched; the unlabelled rate is highest in the early years (all 2017 files, and over 40\% of
the 2020--2022 files), so early-year shares are understated. We did not
independently verify the anchors. Wu et al.~\cite{wu2026esm} instead hand-verify a filtered subset
(177 incidents from the same repository), trading breadth for label precision; our anchor-based
matching covers the full corpus but inherits the authors' original labels, so the two studies
differ in coverage and in the operational meaning of ``price manipulation''.

\paragraph{Static classification.}
Parsing source text without compilation misses inter-contract flows, dynamically constructed
paths and aggregator calls. The audit found 93\% agreement, with both errors under-counting round
trips; the true share is therefore at least as high as reported.

\paragraph{Single coder.}
The 30-case classifier audit and the 47-case coding were performed by one author. Cohen's $\kappa$
from an independent second coder is required before the distribution can be treated as a
measurement result; 21 of the 47 records are medium confidence and 4 are low.
A later consistency pass, re-applying the written rules to the same sources without new evidence,
moved three cases from class B to class C (their prices are moved by donation, not by a swap) and
one from B to D. The headline figures reported here are the post-pass values. That a systematic
re-read changed 12\% of the codes is itself an argument for the second coder, and it is why the
per-case evidence lines are released.

\paragraph{Label-inherited records.}
Three of the 47 records change no price at all and inherit their PM label from the repository's
oracle or inflation keywords (StrategyLlamaLendConvex, Moonwell, EnsoFinance); they are coded C
with an N annotation. Removing them from the subset leaves the headline shares essentially
unchanged: round trips 76/146 = 52.1\%, sunk exits 10/146 = 6.8\%, reserve/accounting 22/146 =
15.1\%.

\paragraph{Execution evidence.}
All three ledgers rest on executions we performed, one per class: Burner on Ethereum, Sharwa on
Arbitrum and SKYDAO on BNB Chain. What the fork work does \emph{not} provide is a systematic
experiment: we did not perturb pool reserves across a range of depths and check whether a sign
flip occurs where a boundary model predicts. Three cases cannot establish that; they establish
only that the three structures are what we say they are, and that the measured settlements match
the cost model we claim applies to each. The perturbation study is the natural next step and the
main open item of this work. It bears on the conditional boundary of Appendix~\ref{app:model},
not on the structural census, which is a static analysis and does not depend on execution.
Fork reproduction requires an archive endpoint and occasionally fails transiently when the
endpoint rate-limits; a retry is part of the released script.

\paragraph{Case-level reconstruction.}
The Burner ledger uses metadata pool depth rather than reserves re-read from the pair, and the
victim's payment is a residual rather than an independent observation. The qualitative conclusion
(the impact is borrowed, not spent) rests on the conservation of the constant-product curve and is
robust; the exact basis points are not.

\section{Conclusion}

Across 901 reproducible DeFi exploits, round trips dominate price-manipulation incidents (51.0\%),
sunk-slippage exits are a minority (6.7\%), and over half of single-leg records move value
without trading at all (53.2\%). Sunk slippage is a real cost in roughly one in fifteen manipulation
incidents, not a general one. The accounting step that actually breaks is fee attribution: whether
the reported profit already contains swaps and flash-loan fees, which a single ``settled'' flag
cannot express.

The immediate implication for tooling is that a feasibility layer should first classify the
execution structure and then apply a class-specific model, and should record fee settlement per
item. The immediate implication for measurement is that corpus-level shares must be reported with
their coding rules and their coder, since the direction of several conclusions reverses depending
on which single-leg cases are miscounted as either round trips or asymmetric exits.

Artifact: classifier, per-case CSV, coding manual and reproduction script are released with this
paper, together with the three fork runs (Ethereum, Arbitrum, BNB Chain) that back the ledgers. A
systematic study that perturbs pool reserves and measures where a predicted boundary actually
flips is left to future work.


\appendix
\section{Coding of the 48 single-leg manipulation files}
\label{app:coding}

Table~\ref{tab:allcoding} lists every record, its chain, its code from
Table~\ref{tab:coding}, the coding confidence ($h$: all three questions answered from explicit
code; $m$: one inferred; $l$: two inferred), and the source lines used as evidence. Line numbers
refer to the proof-of-concept file named in the corpus. The two proofs of the Allbridge incident
are merged into one record (row 13), so the table has 47 rows for 48 files.

\begin{center}
\footnotesize
\begin{longtable}{@{}r p{2.8cm} l c c p{2.6cm} p{4.2cm}@{}}
\caption{Per-case coding with evidence lines}\label{tab:allcoding}\\
\toprule
\# & Case & Chain & Code & Conf. & Lines & Evidence \\
\midrule
\endfirsthead
\toprule
\# & Case & Chain & Code & Conf. & Lines & Evidence \\
\midrule
\endhead
\bottomrule
\endfoot
1 & BelugaDex & Arbitrum & A & h & 82; 122; 142 & flash; swap USDCe$\to$USDT; reverse swap \\
2 & LAURAToken & Ethereum & A & h & 69; 99; 139 & flash; buy LAURA; sell back to WETH \\
3 & MINER & BSC & A & h & 40; 50; 62 & flash; router buy; \texttt{Pair.swap} sell \\
4 & PancakeBunny & BSC & A & h & 103; 105; 141 & \texttt{pairs[0]} both directions; flash \\
5 & PresaleV5 & Ethereum & A & h & 152; 200; 255 & flash; repay; repeat cycle \\
6 & Zoompro & BSC & A & h & 53; 65; 88 & flash borrow; \texttt{buy}; \texttt{sell} \\
7 & BH & BSC & A & m & 67; 86; 134 & oracle flash; dump into pair; router swap back \\
8 & Elephant\_Money & BSC & A & h & 51; 70; 84; 92 & flash swap; buy elephant; sell back same pair \\
9 & Sharwa & Arbitrum & B & h & 168; 188; 228 & inflate NFT collateral; borrow \\
10 & Themis & Arbitrum & B & h & 87; 173; 177 & flash; Balancer swap; borrow \\
11 & Woofi & Arbitrum & B & h & 124; 175; 178 & flash; deposit and borrow; swap out \\
12 & 0x0DEX & Ethereum & B & m & 101; 113; 119 & flash; exploit; swap out \\
13 & Allbridge & BSC & B & m & 64; 113; 117; 77; 89; 97; 105 & flash; deposit/withdraw imbalance; two proofs, one incident \\
14 & CompoundUni & Ethereum & B & m & 55; 75; 78 & flash; borrow UNI; sell \\
15 & DRLVaultV3 & Ethereum & B & m & 78; 108; 135 & flash; vault swap; V3 swap \\
16 & EGD & BSC & B & m & 86; 97; 106 & flash; cross-pool swaps \\
17 & GradientMaker & Ethereum & B & m & 51; 59; 68 & flash; withdraw; swap \\
18 & NeverFall & BSC & B & m & 36; 45; 47 & flash; router buy; victim sell pays out \\
19 & 0vix & Polygon & C & h & 218; 220; 234 & donate GHST to oracle; borrow reserves out \\
20 & Lodestar & Arbitrum & C & h & 199; 241; 252 & mint; donate GLP; borrow \\
21 & PrimeFinance & HyperEVM & C & h & 126; 127; 128 & transfer into pool; take PRFI; borrow \\
22 & DogeAlliance & BSC & C & h & 124; 157; 158 & 1 wei transfer; \texttt{sync} \\
23 & HackDao & BSC & C & h & 29; 39; 41 & \texttt{skim}; \texttt{sync} \\
24 & MetaLend & Ethereum & C & h & 87; 88; 96 & donate ETH; borrow \\
25 & SKYDAO & BSC & C & h & 103; 120; 126 & transfer to pair; drain reserve \\
26 & WiseLending & Ethereum & C & h & 81; 100; 102 & share-price inflation \\
27 & Burntbubba & Ethereum & C & m & 99; 148; 150 & transfer into pairs \\
28 & FireToken & Ethereum & C & m & 45; 66; 103 & flash; withdraw; sell \\
29 & MAMO & BSC & C & m & 43; 59; 68 & flash; pair out; sell \\
30 & MARA & BSC & C & m & 28; 34; 45 & flash; withdraw; pair out \\
31 & MBUToken & BSC & C & m & 49; 56; 64 & deposit; sell \\
32 & SASHAToken & Ethereum & C & m & 78; 88; 100 & buy; transfer into pair; V3 sell \\
33 & Yyds & BSC & C & m & 38; 46; 54 & flash borrow drains pair; claim; transfer into pair \\
34 & ATK & BSC & C & m & 59; 77; 81 & flash borrow drains pair; getPrice distorted; claim \\
35 & VINU & Ethereum & C & h & 49; 60; 66 & seed buy; transfer into pool; sync; swap out \\
36 & SATURN & BSC & C & h & 88; 97; 109 & buy; transfer into pool; \texttt{pair.swap} drain \\
37 & MorphoBlue & Ethereum & C & l & 312; 371; 373 & flash borrow drains PAXG pair; supply collateral; borrow USDC \\
38 & RoulettePotV2 & BSC & C & l & 47; 56; 58; 60 & flash borrow LINK drains pair; finishRound; swapProfitFees \\
39 & StrategyLlama\-LendConvex & Ethereum & C(N) & h & 73; 88; 92 & seed swap; deposit; redeem; no price change (N) \\
40 & Moonwell & Base & C(N) & m & 89; 96; 99 & flash; liquidateBorrow; redeem; no price change (N) \\
41 & YSDAO & BSC & C & h & 105; 108; 111; 119 & buy via pair callback; Staking.sync donate; transfer; sell \\
42 & DLMC & BSC & C & m & 96; 115; 121; 127 & flash; buy; buy; sell at inflated livePrice \\
43 & EnsoFinance & Ethereum & C(N) & m & 140; 151; 160 & buy FARM; deposit; withdrawWETH; no price change (N) \\
44 & DKP & BSC & D & m & 53; 71; 89 & victim pricing not in PoC \\
45 & P719 & BSC & D & m & 91; 127; 142 & multi-contract split, exit not locatable \\
46 & Rodeo & Arbitrum & D & l & 132; 141; 149 & reverse swap and vault deposit (see note) \\
47 & Discover & BSC & D & l & 31; 36; 38 & flash borrow; pledgein; victim mechanism not in PoC; proof does not repay \\
\end{longtable}
\end{center}

\noindent\textbf{Notes.}
\emph{(i)} Rodeo is coded D (undecided) at low confidence: the proof performs a WETH$\to$USDC
swap, a USDC$\to$WETH reverse swap, and then deposits into and borrows against a vault whose share
price depends on the manipulated pair. Both a round trip and an asymmetric exit are present, and
the share of profit attributable to each cannot be read off the source. A second coder should treat
this case as the primary disagreement risk.
\emph{(ii)} Eight files carry code A although the static classifier assigns them to a single leg;
in three of them (BH, PresaleV5, MINER) the return leg runs through a helper contract or a direct
pool call rather than the router used for the entry leg.
\emph{(iii)} Codes C denote reserve or accounting manipulation; the subsequent sale of the
acquired tokens is treated as realisation, not as a price-moving trade. C(N) marks the three
records that change no price at all and inherit their PM label from the repository's oracle or
inflation keywords.

\section{The boundary model and its derivation}
\label{app:model}

This appendix states the model whose scope the census measures. The net-revenue function of
Section~6 is

\begin{equation}
\mathcal{P}_{\text{net}}(P_{\text{gas}},R)
=\mathcal{P}_{\text{gross}}(1-r)-\mathcal{C}_{\text{gas}}(g,P_{\text{gas}})
-(2f+\phi)C-\alpha\frac{C^2}{R+C},
\end{equation}

with $\mathcal{C}_{\text{gas}}(g,P_{\text{gas}})=gP_{\text{gas}}P_{\text{native}}10^{-9}$ and
$\alpha>0$. Write $\kappa_g=gP_{\text{native}}10^{-9}$ and
$\mathcal{B}_0=\mathcal{P}_{\text{gross}}(1-r)-\mathcal{C}_{\text{gas}}-(2f+\phi)C$.

\begin{proposition}[critical gas price]
\label{prop:gas}
For fixed $R>0$, $\mathcal{P}_{\text{net}}$ is strictly decreasing in $P_{\text{gas}}$ on
$[0,\infty)$. If $\mathcal{M}_0(R)\triangleq\mathcal{P}_{\text{gross}}(1-r)-(2fC+\alpha C^2/(R+C))-C\phi>0$,
there is a unique positive root
$P_{\text{gas}}^{*}(R)=\mathcal{M}_0(R)/\kappa_g$, and the attack is feasible at this depth if and
only if $P_{\text{gas}}<P_{\text{gas}}^{*}(R)$; if $\mathcal{M}_0(R)\le 0$, no non-negative gas
price makes it feasible.
\end{proposition}

\begin{proof}
$\mathcal{P}_{\text{net}}=\mathcal{M}_0(R)-\kappa_g P_{\text{gas}}$ with $\kappa_g>0$, so the
derivative is $-\kappa_g$; the function is continuous, equals $\mathcal{M}_0(R)$ at $P_{\text{gas}}=0$
and tends to $-\infty$. Strict monotonicity plus the intermediate value theorem give existence and
uniqueness of a zero when $\mathcal{M}_0(R)>0$, and its absence otherwise.
\end{proof}

\begin{proposition}[critical pool depth]
\label{prop:tvl}
For fixed $P_{\text{gas}}\ge 0$, $\mathcal{P}_{\text{net}}$ is strictly increasing and strictly
concave in $R$ on $(0,\infty)$, with
$\partial_R\mathcal{P}_{\text{net}}=\alpha C^2/(R+C)^2>0$ and
$\partial^2_R\mathcal{P}_{\text{net}}=-2\alpha C^2/(R+C)^3<0$. If
$0<\mathcal{B}_0<\alpha C$, the unique positive root is
\begin{equation}
R^{*}=\frac{\alpha C^2}{\mathcal{B}_0}-C,\qquad \text{TVL}^{*}=2R^{*};
\end{equation}
if $\mathcal{B}_0\le 0$ the attack is infeasible at every depth, and if
$\mathcal{B}_0\ge\alpha C$ it is feasible at every depth.
\end{proposition}

\begin{proof}
The derivatives follow by differentiation, with $C>0$ and $\alpha>0$. As $R\to 0^{+}$,
$\mathcal{P}_{\text{net}}\to\mathcal{B}_0-\alpha C$; as $R\to\infty$ it tends to
$\mathcal{B}_0$. Strict monotonicity and the intermediate value theorem give a unique root exactly
when $0<\mathcal{B}_0<\alpha C$, and solving $\mathcal{P}_{\text{net}}=0$ yields the expression.
\end{proof}

\begin{corollary}[sharpness]
The normalised slope at $R=R^{*}$, as a function of $k=\text{TVL}/\text{TVL}^{*}$, is
$s=\alpha C\,u(1-u)/\mathcal{P}_{\text{gross}}$ with $u=\mathcal{B}_0/(\alpha C)$; it is maximal at
$u=1/2$ and vanishes at both ends.
\end{corollary}

\begin{proof}
At $R=R^{*}$ one has $R^{*}+C=\alpha C^2/\mathcal{B}_0$, hence
$\partial_R\mathcal{P}_{\text{net}}=\mathcal{B}_0^2/(\alpha C^2)=u^2\alpha$ and
$\mathrm{d}R/\mathrm{d}k=R^{*}=C(1-u)/u$, so
$\mathrm{d}\mathcal{P}_{\text{net}}/\mathrm{d}k=\alpha C u(1-u)$; divide by
$\mathcal{P}_{\text{gross}}$ and maximise $u(1-u)$.
\end{proof}

\noindent\textbf{What the propositions do not say.} All three results are conditional on
$\alpha>0$ and on $\mathcal{P}_{\text{gross}}$ not depending on $R$. The first fails in class A,
where the round trip returns the position and $\alpha=0$: the model then has no root in $R$ at all,
and feasibility is decided by fees alone. The second fails in class A as well, because the victim's
payment scales with the distortion $C/(R+C)$ and net revenue decreases in $R$, which converts the
lower bound into an upper one. In class C there is no exchange at all, so no term of this form can
be defined. The propositions are therefore correct but narrow, and the census in Section~4 is what
establishes how narrow.


\begin{thebibliography}{99}
\small
\setlength{\itemsep}{0.25em}

\bibitem{zhou2023sok} Zhou L, Xiong X, Ernstberger J, et al. SoK: Decentralized Finance (DeFi) Attacks[C]//IEEE S\&P. 2023: 2444--2461.
\bibitem{werner2022sok} Werner S, Perez D, Gudgeon L, et al. SoK: Decentralized Finance (DeFi)[C]//ACM AFT. 2022: 30--46.
\bibitem{flashloans} Qin K, Zhou L, Livshits B, et al. Attacking the DeFi Ecosystem with Flash Loans for Fun and Profit[C]//FC. 2021: 3--32.
\bibitem{foray} Wen H, Liu H, Song J, et al. FORAY: Towards Effective Attack Synthesis against Deep Logical Vulnerabilities in DeFi Protocols[C]//ACM CCS. 2024: 1001--1015.
\bibitem{defiposer} Zhou L, Qin K, Cully A, et al. On the Just-In-Time Discovery of Profit-Generating Transactions in DeFi Protocols[C]//IEEE S\&P. 2021: 919--936.
\bibitem{ityfuzz} Shou C, Tan S, Sen K. ItyFuzz: Snapshot-Based Fuzzer for Smart Contract[C]//ACM ISSTA. 2023: 322--333.
\bibitem{a1} Gervais A, Zhou L. AI Agent Smart Contract Exploit Generation[EB/OL]. arXiv:2507.05558, 2026.
\bibitem{defiranger} Wu S, Yu Z, Wang D, et al. DeFiRanger: Detecting DeFi Price Manipulation Attacks[J]. IEEE TDSC, 2024. DOI: 10.1109/TDSC.2023.3346888.
\bibitem{pmdetector} Liu L, Zhang W, Wei L, et al. LLM-Powered Detection of Price Manipulation in DeFi[EB/OL]. arXiv:2510.21272, 2025.
\bibitem{v2e} Zhang J, Nan Y, Ning K, et al. V2E: Validating Smart Contract Vulnerabilities through Profit-driven Exploit Generation and Execution[J]. Proc. ACM Softw. Eng., 2026, 3(FSE): Article FSE101.
\bibitem{ethor} Schneidewind C, Grishchenko I, Scherer M, et al. eThor: Practical and Provably Sound Static Analysis of Ethereum Smart Contracts[C]//ACM CCS. 2020: 621--640.
\bibitem{over} Deng X, Beillahi S M, Minwalla C, et al. Safeguarding DeFi Smart Contracts against Oracle Deviations[C]//ICSE. 2024.
\bibitem{flashboys} Daian P, Goldfeder S, Kell T, et al. Flash Boys 2.0[C]//IEEE S\&P. 2020: 910--927.
\bibitem{qin2022mev} Qin K, Zhou L, Gervais A. Quantifying Blockchain Extractable Value: How Dark is the Forest?[C]//IEEE S\&P. 2022: 198--214.
\bibitem{cff} Babel K, Daian P, Kelkar M, et al. Clockwork Finance: Automated Analysis of Economic Security in Smart Contracts[C]//IEEE S\&P. 2023.
\bibitem{angeris2020oracles} Angeris G, Chitra T. Improved Price Oracles: Constant Function Market Makers[C]//ACM AFT. 2020: 80--91.
\bibitem{angeris2021uniswap} Angeris G, Kao H T, Chiang R, et al. An Analysis of Uniswap Markets[EB/OL]. arXiv:1911.03380, 2021.
\bibitem{xu2023dex} Xu J, Paruch K, Cousaert S, et al. SoK: Decentralized Exchanges (DEX) with Automated Market Maker (AMM) Protocols[J]. ACM Computing Surveys, 2023, 55(11).
\bibitem{qin2023imitation} Qin K, Chaliasos S, Zhou L, et al. The Blockchain Imitation Game[C]//USENIX Security. 2023.
\bibitem{gudgeon2020defi} Gudgeon L, Werner S, Perez D, et al. DeFi Protocols for Loanable Funds[C]//ACM AFT. 2020: 92--112.
\bibitem{wu2026esm} Wu H, Wang H, Luo X, et al. Price is Only the Symptom: Understanding and Analyzing Economic-State Manipulation Vulnerabilities in DeFi[EB/OL]. arXiv:2611.00587, 2026. (accepted by ACM, 2026)

\end{thebibliography}
\end{document}